\documentclass[a4paper,11pt]{article}

\usepackage[english]{babel}
\usepackage[a4paper]{geometry}
\usepackage{amsmath, amsfonts, systeme, mathtools}
\usepackage{setspace}
\usepackage{graphicx}
\usepackage[colorlinks=true, citecolor=blue, linkcolor=blue]{hyperref}
\usepackage[biblabel]{cite}

\usepackage{algorithm, algorithmic}
\usepackage{enumitem}
\usepackage{listings}
\usepackage{xcolor}
\usepackage{sectsty}
\usepackage{notoccite}

\usepackage{libertine}

\usepackage{titlesec}
\titlespacing{\section}{0pt}{0pt}{0pt}
\usepackage{caption}
\begin{document}

\begin{flushleft}
	\vspace{-2cm}
	{\Large \textsf{A Generalized Formulation for Accurate and Robust Determination of Soil Shear Strength from Triaxial Tests}}\\[1em]
	{\small \textsf{ALTAMIRANO MUÑIZ EMILIO FERNANDO}}\\
	{\small \textsf{mail: amz@gmx.cn}}
	\vspace{0cm}
\end{flushleft}

\begin{flushleft}
	{\small \textbf{Abstract}}\\
	{\footnotesize 	 This work presents an extended formulation of the Least Squares with Virtual Displacements (LSVD) method for estimating shear strength parameters from multiple soil samples under varying resistance conditions including cohesionless, frictional, and mixed types. LSVD is designed to identify a common tangent across $n$ Mohr circles, even in the presence of measurement errors that render an exact solution infeasible. Beyond its original linear formulation, we introduce generalized LSVD variants like logarithmic, parabolic, polynomial, power law and generalized forms allowing the method to adapt to diverse failure envelope shapes observed in geotechnical materials.}\\
	
	{\footnotesize We benchmark these variants against established approaches such as the $p$-$q$ method and CTPAC, analyzing performance under synthetic noise to simulate measurement uncertainty. This provides a comparative framework to assess each method’s robustness, especially considering their differing selections of representative points on the Mohr circles. The results highlight LSVD's flexibility and reliability in modeling complex soil behavior and suggest its potential as a versatile tool for geomechanical analysis.}
\end{flushleft}

\section{Introduction}
This work revisits and extends the Least Squares with Virtual Displacements (LSVD) method as a general framework for obtaining approximate failure envelopes from $n$ soil samples, even in the presence of imperfect measurements. LSVD formulates the problem as an optimization task, minimizing a squared-error loss function defined over points generated through virtual displacements. Unlike standard regression approaches, LSVD does not operate on fixed datapoints; instead, the datapoints evolve with each trial solution. This feature enables a flexible formulation that is consistent with tangent failure theory. Although the Mohr–Coulomb failure criterion theoretically allows the construction of a common tangent to a set of Mohr circles, this procedure is rarely straightforward in practice. Experimental variability and measurement error in laboratory data often preclude the existence of a single line tangential to all samples, rendering an analytical solution unattainable.\\

Several methods have been developed to approximate the failure envelope. Among them, the \textit{p-q} method is widely used due to its simplicity and relatively accurate estimates. It performs a linear regression on the centers and radii of the Mohr circles, effectively assuming that failure occurs near the center. However, it tends to underestimate shear strength and does not directly adhere to the Mohr–Coulomb failure theory. Another technique, CTPAC, computes common tangents between adjacent circles and fits a regression line through the derived points. While it may recover the exact solution when it exists, it lacks robustness in the presence of noise or nonlinearity. In this work, we compare these two approaches, emphasizing that both rely solely on geometrical considerations, unlike other methods such as Hoek–Brown, which incorporate additional parameters including the Geological Strength Index (GSI), stress conditions, and other material properties.\\

Beyond the original linear formulation, we propose and implement several generalized LSVD variants such as logarithmic, parabolic, polynomic, and power-law envelopes each adapting to different classes of soil behavior, including materials that exhibit strength degradation at high stress levels \cite{trist}. These variants preserve the core least-squares principle while extending the modeling capacity of LSVD. The performance of each formulation is evaluated and compared against classical methods under simulated noise conditions, providing a robust framework for analyzing soils with complex or uncertain failure patterns.\\
\section{Formulation}
To apply the least squares approach, it is first necessary to define a loss function, given in equation \ref{loss}:

\begin{equation}
	\ell(\tau, \hat{s}) = \sum_{i} \left( \tau_i - \hat{s}_i \right)^2
	\label{loss}
\end{equation}

Here, $\hat{s}_i$ represents the estimated shear failure envelope, while $\tau_i$ denotes the actual shear stress at failure. In practice, $\tau_i$ is not directly known; although the equations of the Mohr circles can be derived from experimental data, the precise failure points on those circles remain unidentified.

To address this, the method introduces the concept of \textit{virtual displacements}. This reformulates the problem as one of determining a straight line that is either tangent to all Mohr circles or, alternatively, minimizes the square difference between a \textit{virtual tangent} and the corresponding semicircle.

In this approach, the loss function compares the point $\prescript{\theta}{}{\sigma_i}$ obtained from the estimated envelope $\hat{s}_i=\beta_1 \prescript{\theta}{}{\sigma_i} + \beta_0$ with the point on the semicircle $\tau_i= \sqrt{r_i^2-(\prescript{\theta}{}{\sigma_i}-\lambda_i)^2}$. The value of $\prescript{\theta}{}{\sigma_i}$ is determined by solving equation \ref{ends}, which identifies the location at which the line would be tangent to the circle.

\begin{figure}[t!]
	\begin{tabular}{c c}
		\includegraphics[width= 0.3\linewidth]{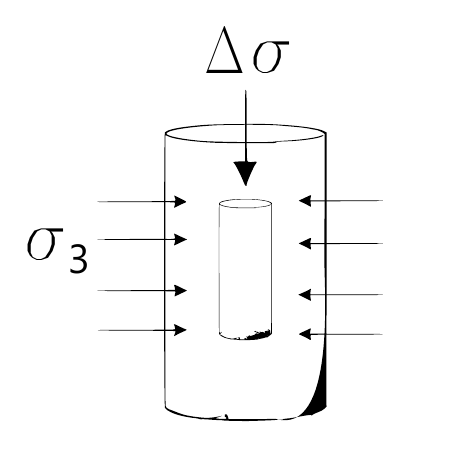} & \includegraphics[width= 0.5\linewidth]{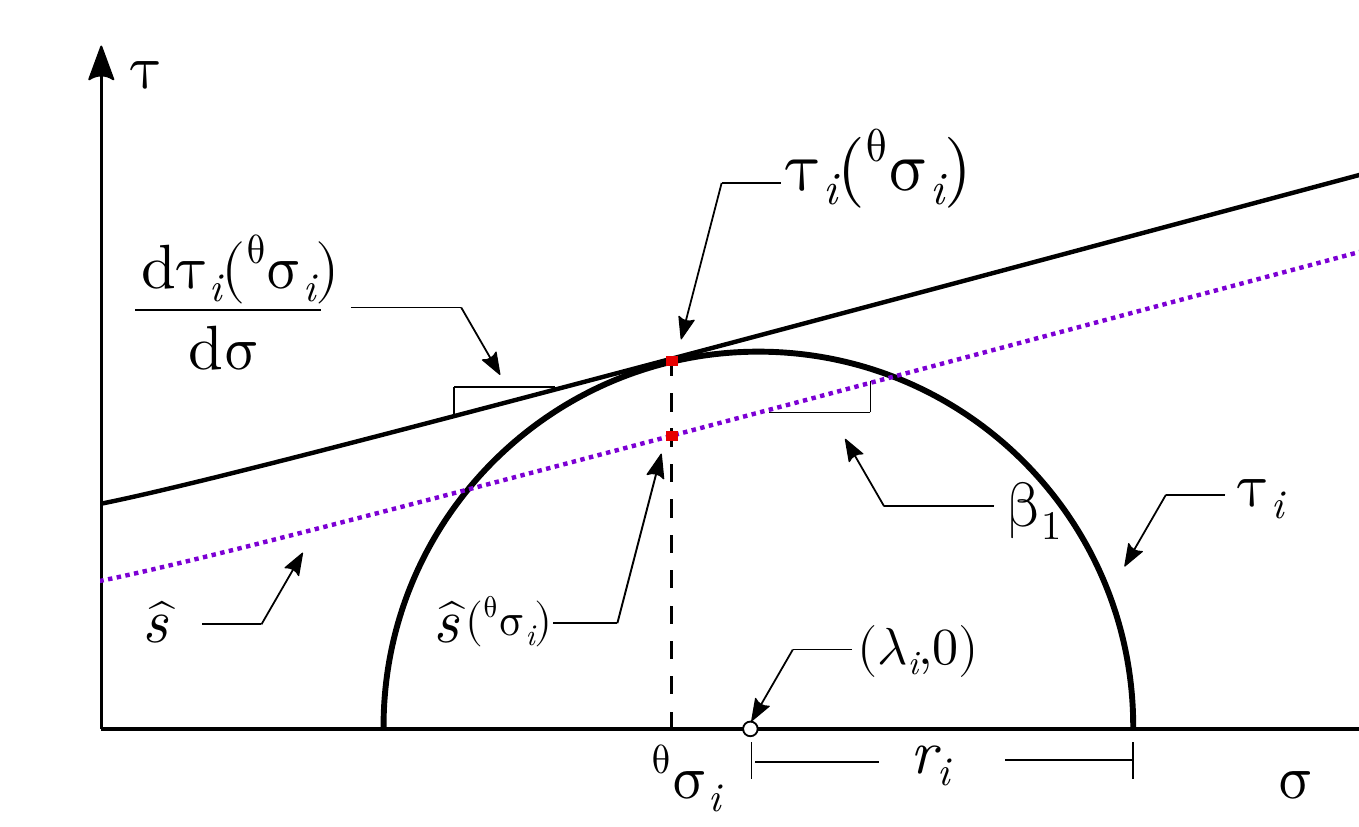} \\
	\end{tabular}
	\caption{Scheme of stresses in a soil sample (left) and the virtual displacement by the $\sigma_3<\prescript{\theta}{}{\sigma_i}<\sigma_1$ to compute least square error (right).}
	\label{scheme}
\end{figure}

\begin{equation}
	\frac{d\tau_i}{d\sigma} = \frac{d\hat{s}_i}{d\sigma} \\
	\label{ends}
\end{equation}
\subsection{Linear and Logarithmic Failure Envelope}
This is the case of the Mohr-Coulomb linear criteria, so the estimation $\hat{s} = \beta_1 \sigma + \beta_0$ is the failure envelope, where the bias $\beta_0$ and $\beta_1$ happens to be the cohesion and friction angle's tangent, respectively. Then $\hat{s}'$ is $\beta_1$ and the point where the virtual displacement will happen will satisfy equation \ref{mm1}.
	\begin{equation}
		\frac{d\tau_i}{d\sigma}=\beta_1\\
		\label{mm1}
	\end{equation}
The solution to equation \ref{mm1} is the equation \ref{sol1}. It is only taken the solution that makes sense with the failure envelopes. This solution requires $\lambda_i$ to be positive, a harmless assumption for most of laboratory tests. Although the final equation to find $\beta_1$ can take circles on the negative side.
	\begin{equation}
		\prescript{\theta}{}{\sigma_i} = \lambda_i - \frac{\beta_1 r_i}{\sqrt{\beta^2_1 + 1 }}
		\label{sol1}
	\end{equation}
Thus, the loss function is
\begin{equation}
		\ell(\tau, \hat{s}) = \sum_{i} \left\{ \sqrt{r_i^2 - \left(- \frac{\beta_1 r_i}{\sqrt{\beta_1^2 + 1}}\right)^2} \right. 
		\left. - \beta_1 \left(\lambda_i - \frac{\beta_1 r_i}{\sqrt{\beta_1^2 + 1}}\right) - \beta_0 \right\}^2
		\label{sol2}
\end{equation}

If $n=1$ the equation \eqref{sol2} will always sum zero (explanation on appendix \ref{appendixc}). If that is the case then there are no reference points and assumptions are made like a cohesionless condition, so it is recommended to opt for analytical solutions that can be found on soil mechanics textbooks. An analytical solution is the equation $\varphi = \arctan (r^2/\lambda \sqrt{r^2 - r^4/\lambda^2})$ for the friction angle.\\

The logarithmic equation has the form $\hat{s} = \beta_1 \ln(\sigma + \beta_2) + \beta_0$, and the function to compute the $\bar{\beta}$ with the logarithmic form is equation \ref{loga_fu}.
\begin{equation}
	\bar{\beta} = \arg \min_{\beta_0,\beta_1, \beta_2}
	\left(\sum_{i} \Bigl\lbrace
	\sqrt{r_i^2 - (\prescript{\theta}{}{\sigma_i}-\lambda_i) ^2} 
	-\beta_1 \ln \left( \prescript{\theta}{}{\sigma_i} + \beta_2 \right) - \beta_0 \Bigr\rbrace ^2\right)
	\label{loga_fu}
\end{equation}\\
A sharp reader will notice that the solution to \eqref{sol2} is essentially the linear regression formula modified but with $\prescript{\theta}{}{\sigma_i}$ plugged in. As it was mentioned in the introduction, this method uses the square loss and the virtual displacements (VD) to formulate a function to optimize but the VD's are not actual fixed points, so what it is happening is that the VD will depend also on the coefficients, i.e., the $x$ \textit{exes} ($\prescript{\theta}{}{\sigma}$ for us) depend on the $\beta_1$, then it not a regression problem.
	
\subsection{Parabolic Failure Envelope}
This form is good for a lot of samples and when it is desired to have a fixed exponent, the general form is discussed in the following section, but these are only selected functions, and LSVD can take any form.
	\begin{equation}
		\hat{s}=\beta_1 \sqrt{\sigma} + \beta_0
		\label{parabolic}
	\end{equation}
Therefore the VD will occur on the solution of equation \eqref{parabolic_VD}.
	\begin{equation}
		0 = \frac{-\prescript{\theta}{}{\sigma_i} + \lambda_i}{\sqrt{r^2_i - (\prescript{\theta}{}{\sigma_i} - \lambda_i)^2}} - 
		\dfrac{\beta_1}{2\sqrt{\prescript{\theta}{}{\sigma_i}}}
		\label{parabolic_VD}
	\end{equation}
\subsection{Power Failure Envelope}
A more general form of the parabolic function is the power equation with the form of equation \eqref{power}, sometimes the $\beta_0$ (bias) is not included for this form, so let us call the equation \textit{full power} equation. In order to have a semi parabola the constraint $0<\beta_3<1$ is sought. The coefficient $\beta_2$ will move the function horizontally and the bias $\beta_0$ will move the function vertically, so it is reasonable that various combinations of coefficients will suffice the desired result.
	\begin{equation}
		\hat{s}=\beta_1 \left( \sigma + \beta_2 \right) ^{\beta_3} + \beta_0
		\label{power}
	\end{equation}
Therefore the virtual displacement will occur on the solution of equation \eqref{powe_vd}
	\begin{equation}
		0 = \frac{-\prescript{\theta}{}{\sigma_i} + \lambda_i}{\sqrt{r^2_i - (\prescript{\theta}{}{\sigma_i} - \lambda_i)^2}} - \beta_1 \beta_3 \left( \prescript{\theta}{}{\sigma_i} + \beta_2 \right) ^ {\beta_3-1}
		\label{powe_vd}
	\end{equation}
And the minimization is formulated as shown in equation \eqref{power_min}, but this problem has various local minima, so depending in the optimization method and the initial guesses will be the solution, but most of the solutions should be practical (this is mere speculation from the experience of this work).
	\begin{equation}
		\bar{\beta} = \arg \min_{\beta_0,\hdots, \beta_3}
		\ell(\tau, \hat{s}) = \sum_{i} \Bigl\lbrace
		\sqrt{r_i^2 - (\prescript{\theta}{}{\sigma_i}-\lambda_i) ^2} 
		-\beta_1 \left( \prescript{\theta}{}{\sigma_i} + \beta_2 \right) ^{\beta_3} - \beta_0 \Bigr\rbrace ^2
		\label{power_min}
	\end{equation}
\subsection{Polynomial Envelope}
\begin{figure}[!h]
	\centering
	\includegraphics[width= 300pt]{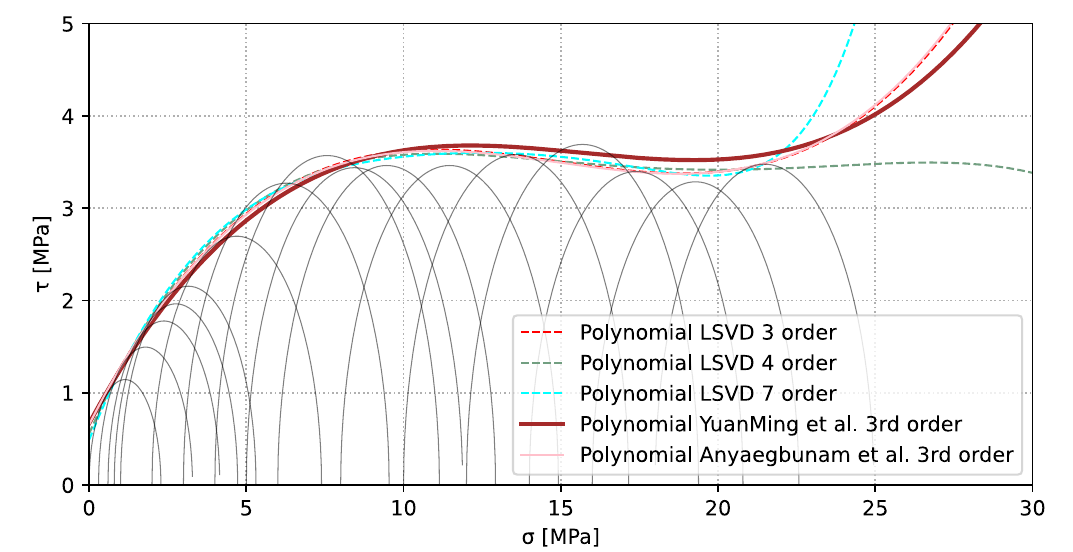}
	\caption{Failure envelopes with the polynomial LSVD and the results of \cite{calib} and \cite{yuanming}}
	\label{poly_res}
\end{figure}

This form looks like a linear model at first glance but it is not really because $\tau_i$ are not known and will be computed on the process. It can be viewed as the general form \eqref{pol_s} of the linear envelope and the linear form being the particular case when $p = 1$ and $p$ being the degree of the polynomial.
	\begin{equation}
		\hat{s} = \sum_{j=0}^{p} \beta_j \sigma ^j
		\label{pol_s}
	\end{equation}
The virtual displacement will be taken from the solution of equation \eqref{pol_vd}
	\begin{equation}
		0 = \frac{-\prescript{\theta}{}{\sigma_i} + \lambda_i}{\sqrt{r^2_i - (\prescript{\theta}{}{\sigma_i} - \lambda_i)^2}} - 
		\sum_{j=1}^{p} (j) \beta_j \left(\prescript{\theta}{}{\sigma_i} \right)  ^ {j-1}
		\label{pol_vd}
	\end{equation}
A comment on this formulation is that the VD point $\prescript{\theta}{}{\sigma_i}$ can lie within all the domain of each semi-circle $(\prescript{i}{}{\sigma_3},\prescript{i}{}{\sigma_1})$ as opposed to the previously discussed methods. But it has the limitation of the polynomial regression which is overfitting. From the data and results of \cite{yuanming} that is on table 1 the LSVD returns the results shown in figure \ref{poly_res}.\\
\begin{table}
	\label{table_yua}
	\caption{Test data from YuanMing \cite{yuanming}}
	\centering
	\begin{tabular}{ c c c c c c c c c}
	\hline		
	Stress	&	& & &[$MPa$]\\	
	\hline																
	$\sigma_3$	&	0	&	0.3	&	0.6	&	0.8	&	1	&	2	&	3	&	4	\\
	$\sigma_1$	&	2.285	&	3.289	&	4.155	&	4.726	&	5.308	&	7.392	&	9.541	&	11.14	\\
	\hline																
	$\sigma_3$	&	5	&	6	&	8	&	10	&	12	&	14	&	16	&	18	\\
	$\sigma_1$	&	11.877	&	12.924	&	14.924	&	17.161	&	19.381	&	20.795	&	22.571	&	24.953	\\
	\hline																
	\end{tabular}
\end{table}
The equation calculated by YuanMing \cite{yuanming} is $\hat{s} = 0.6667 + 0.629 \sigma-0.0423 \sigma ^ 2 + 0.0009\sigma ^ 3$, from Anyaegbunam et al. \cite{calib} they got $\hat{s}= 0.6549 + 0.6690 \sigma - 0.04815 \sigma ^2 + 0.0010773 \sigma ^3$ and by LSVD is $\hat{s} = 0.652918997 + 0.671677577 \sigma  -0.048268340 \sigma ^ 2 + 0.001076909\sigma ^ 3$, as it is shown in figure \ref{poly_res}.\\

So it can be seen that it overfits the data and definitely does not behaves well with high orders nor high pressures, and the constrained least-squares ought to be used with the bounds needed by the specific case if that is desired. 
\subsection{Other Non Linear Failure Envelope}
The linear, \textit{full power} and logarithmic forms should be able to cover a wide range of non linear failure envelopes but LSVD is more flexible than that, the background is already stablished for any kind of function. The algorithm is \ref{lsvd_algo}. Doing step 3 and 4 might be only done numerically on algorithm \ref{lsvd_algo}, so that process should be within the optimization process and solve that part first, in the first iteration the calculation will be done with the initial guess of $\beta_0, \hdots, \beta_n$.\\
	\begin{figure}[h!]
		\centering
		\begin{algorithm}[H]
			\caption{General formulation of LSVD}
			\label{lsvd_algo}
			\begin{algorithmic}[1]
				\renewcommand{\algorithmicrequire}{\textbf{Input:}}
				\renewcommand{\algorithmicensure}{\textbf{Output:}}
				\REQUIRE $r_i$, $\lambda_i$ or $\prescript{i}{}{\sigma_3}$, $\prescript{i}{}{\sigma_1}$
				\STATE Define $\hat{s}(\sigma, \beta_0, \hdots, \beta_n)$
				\STATE Get $\frac{\partial\hat{s}}{\partial\sigma}$
				\STATE Solve $\frac{d\tau}{d\sigma}-\frac{\partial\hat{s}}{\partial\sigma}= 0$ for $\prescript{\theta}{}{\sigma}$
				\STATE Substitute $\prescript{\theta}{}{\sigma_i}$ on $\ell_i(\tau_i,\hat{s}_i)$
				\STATE $\beta_0,\hdots, \beta_n \leftarrow \arg\min_{\beta_0,\hdots,\beta_n} {\sum_{i}\ell_i}$
				\RETURN $\beta_0, \hdots, \beta_n$
			\end{algorithmic} 
		\end{algorithm}
	\end{figure}
Some important notes to remark are that depending of the function it could happen that there are various solutions to the virtual displacement \eqref{ends}, knowing that the solution is real and lies in the first half of the circle helps to discriminate solutions, but it could not be the case, bear that in mind. It is helpful to plot the derivatives of the circles with the derivative of $\hat{s}$ to see if there is a solution. Other note is that minimizing the loss/objective function can quickly become a complex problem with many local minima, so it is up to the minimization method and how the problem is approached.\\

Other strategies to delimit the problem is to fix one coefficient or remove it, for instance, with the \textit{full power} form $\beta_0$ could be taken out which makes sense for a cohesionless soil or fix the $\beta_3$ and compute the other coefficients.

\section{Results}
The sample's data is shown in table \ref{tabul}. Given that there is no exact solution to those soil samples, and each methods compares from different points, then it is suitable to analyze their variability and confidence intervals.\\
	\begin{table}[h!]
		\centering
		\begin{tabular}{ c c c c c c c c}
			\hline
			\textbf{Sample}  \\
			\hline
			 1 & $\sigma_3$  & 2.280 & 4.550 & 8.830 & 17.390 & 34.700 & 69.100    	 \\
			 & $\sigma_1$ & 19.250 & 28.010 & 52.370 & 82.590 & 157.200 & 288.600 	 \\
      \hline
		\end{tabular}
	
	\caption{Failure pressures of samples [$MPa$]}
	\label{tabul}
	\end{table}
Laboratory measurements will carry errors due to the equipment, enviroment or other variables, so the samples' data were put under random noise of 10\%, $\epsilon \sim \text{Unif}[-0.1, 0.1]$. The random noise is also used to analyze the response of the LSVD with sample errors, then an assumption is that the measured data has no errors, and describe the real circles.\\

\begin{figure}[t!]
		\begin{tabular}{c c}
			\includegraphics[width= 0.5\linewidth]{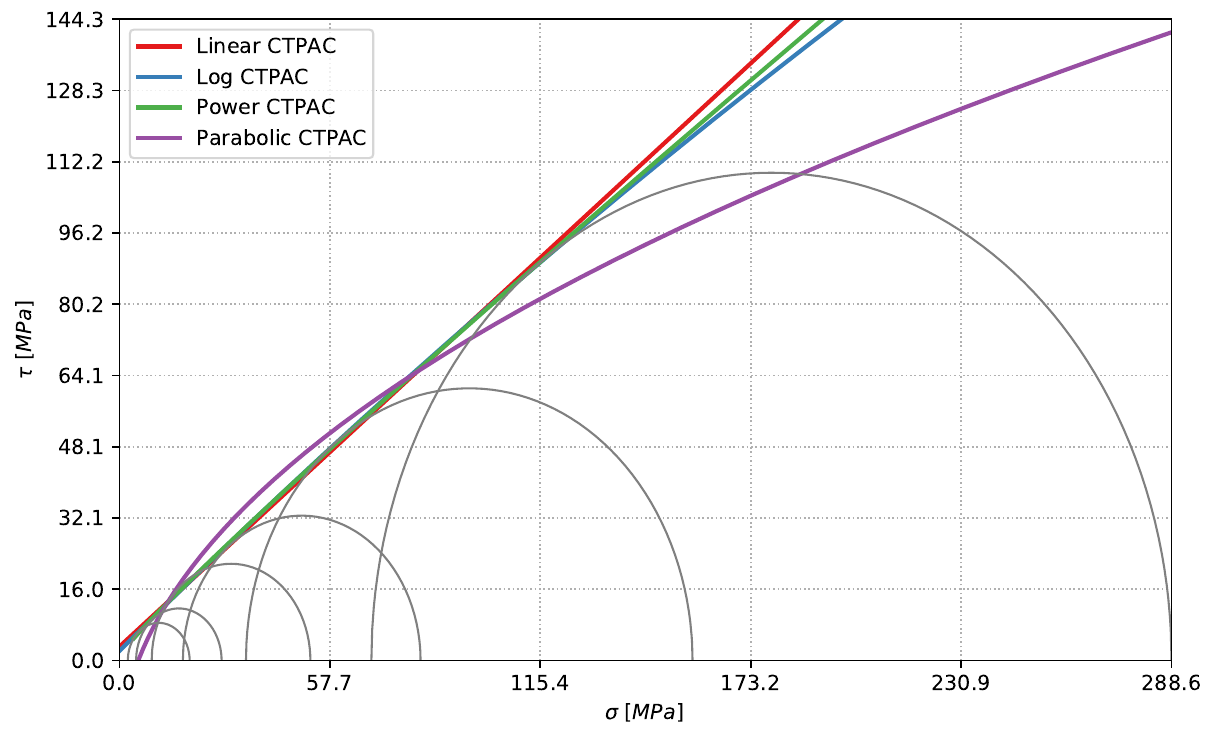} & \includegraphics[width= 0.5\linewidth]{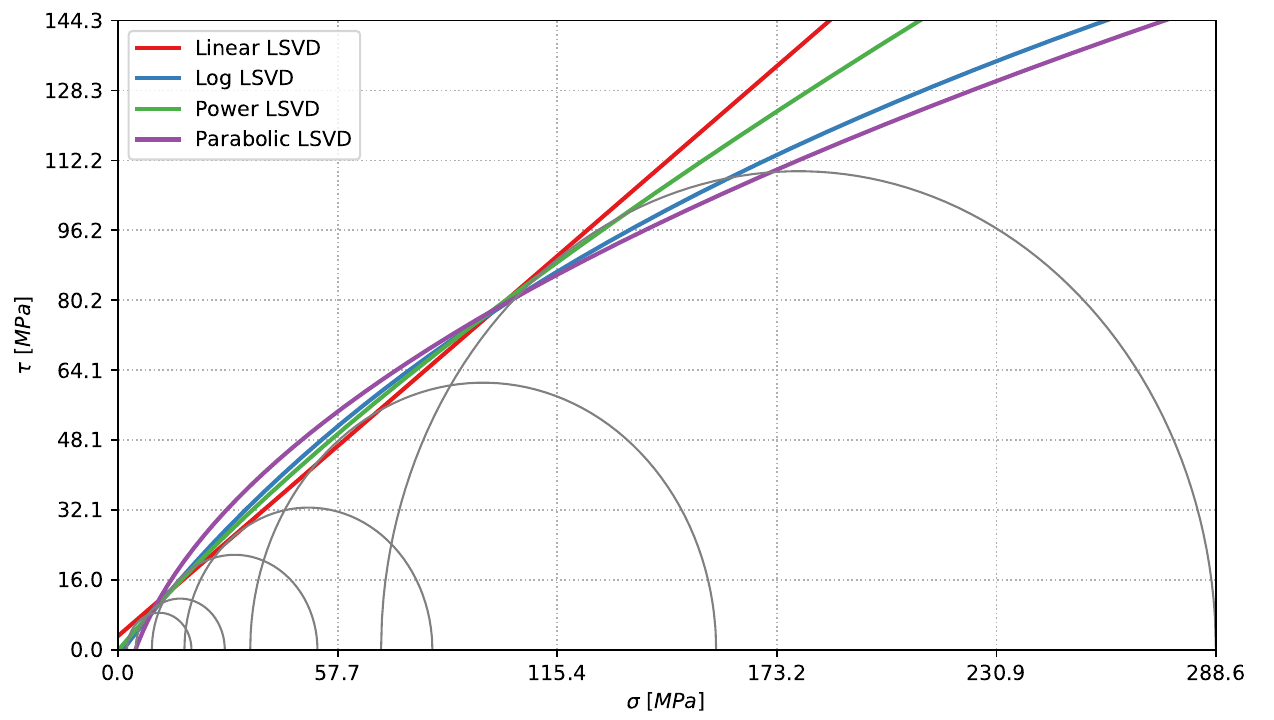} \\
			\includegraphics[width= 0.5\linewidth]{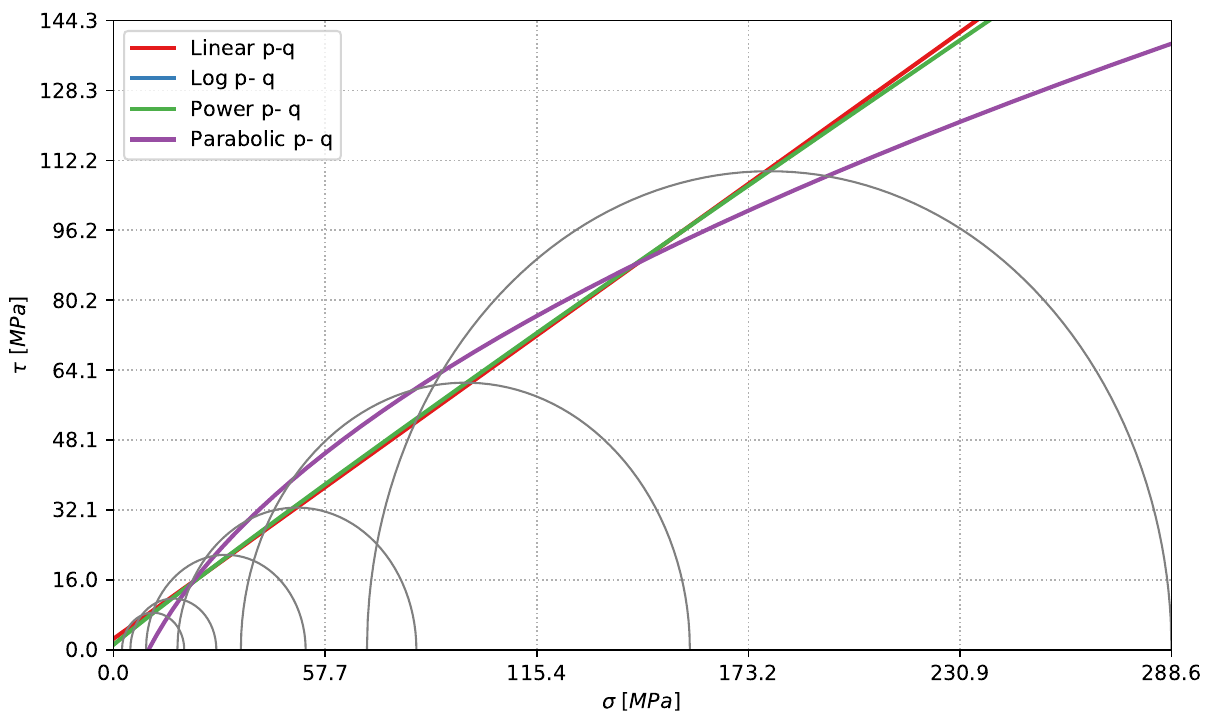} &  \\
		\end{tabular}
		\caption{Failure envelopes computed with different methods. Top-left: CTPAC method. Top-right: LSVD method. Bottom-left: p-q method}
		\label{results_methods}
	\end{figure}

\begin{figure}[t!]
		\begin{tabular}{c c}
			\includegraphics[width= 0.5\linewidth]{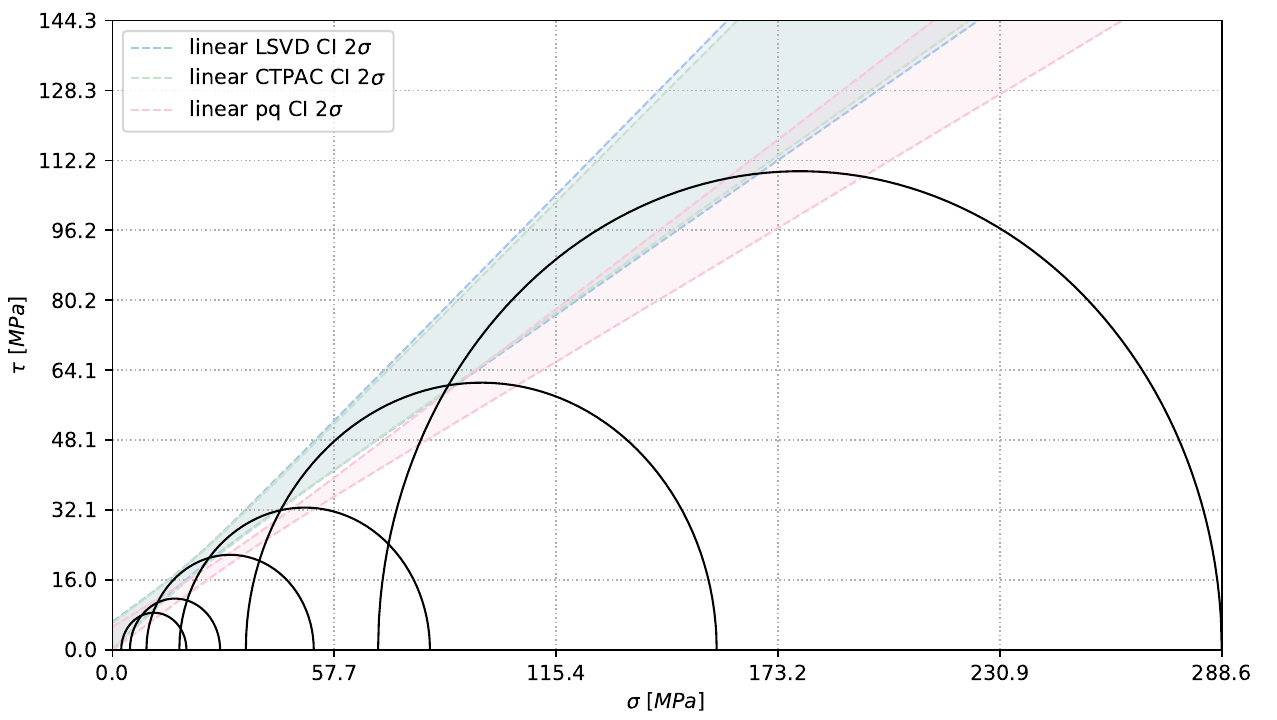} & \includegraphics[width= 0.5\linewidth]{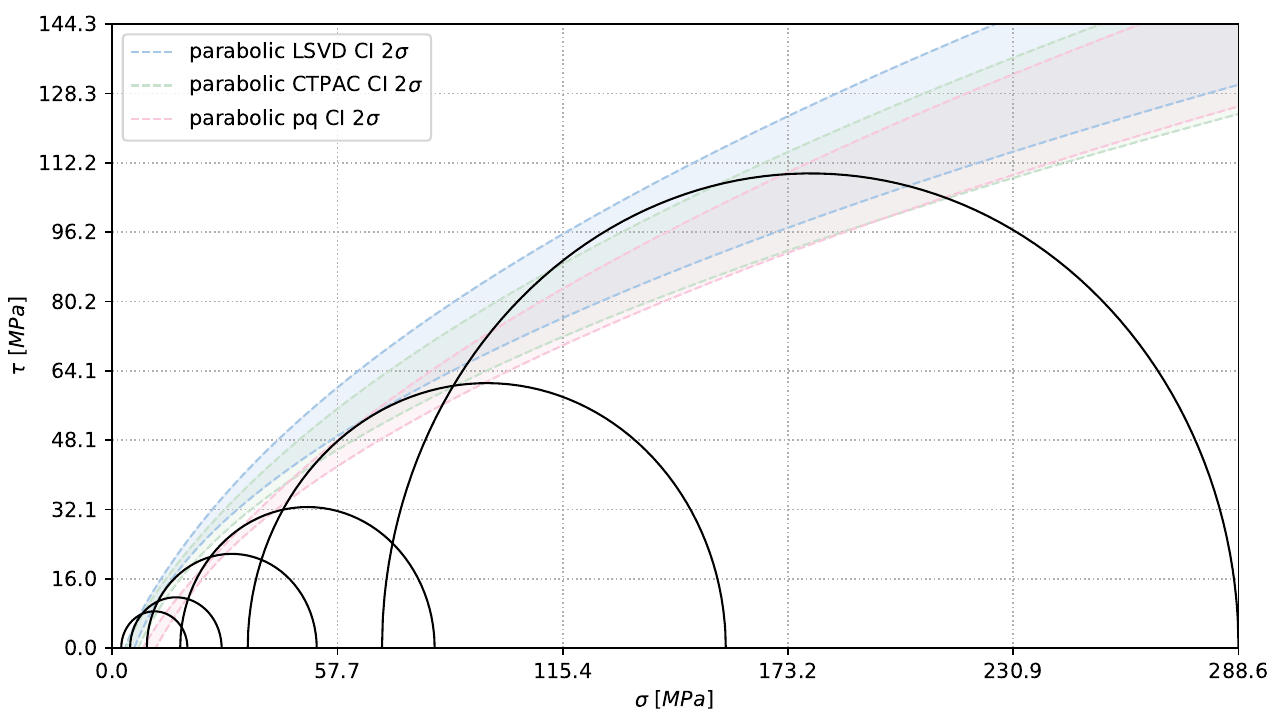} \\
			\includegraphics[width= 0.5\linewidth]{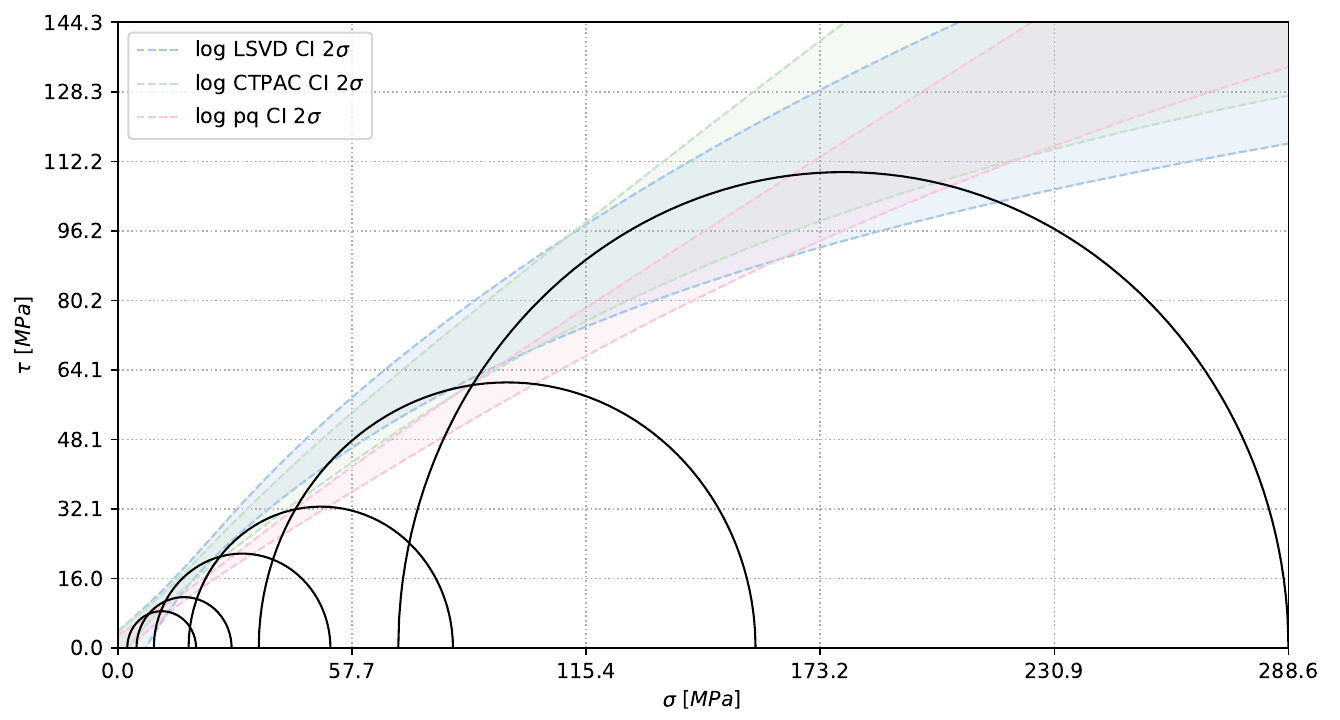} & \includegraphics[width= 0.5\linewidth]{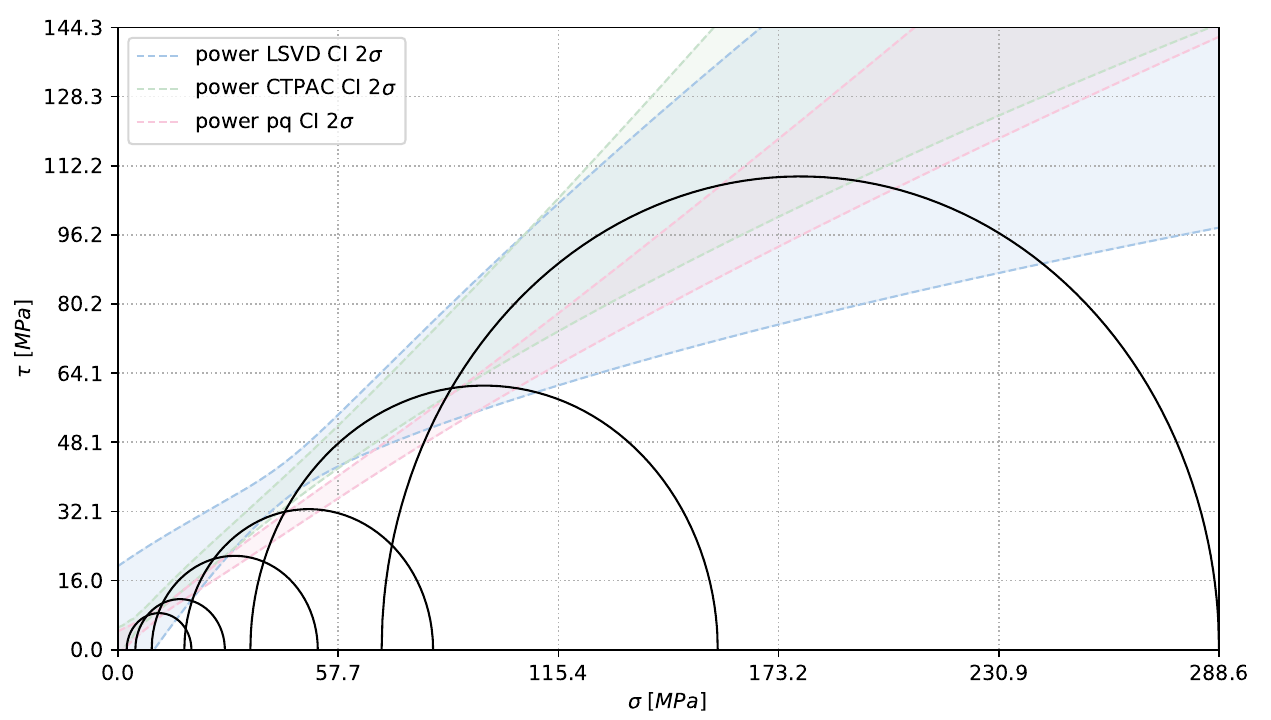} \\
		\end{tabular}
		\caption{Variability computed with different methods. Top-left: linear methods. Top-right: parabolic methods. Bottom-left: logarithmic methods. Bottom-right: power methods.}
		\label{results_methods_CI}
	\end{figure}

Figure \ref{results_methods} show the sample with their failure envelopes corresponding to their methods. Figure \ref{results_methods_CI} are the confidence intervals with two standard deviations under random noise.\\
\newpage
\section{Discussion}
  From an algorithmic and mathematical perspective, the LSVD approach could, in principle, be reformulated to compute perpendicular displacements from the closest parallel line rather than vertical virtual displacements. However, the method deliberately avoids such additional calculations, relying instead on vertical displacement while still ensuring that tangency can be identified.\\
  
  The LSVD method explicitly enforces tangency between the candidate envelope and the Mohr circles, incorporating it into the optimization process while accounting for potential measurement errors. In contrast, the p–q method ignores tangency as a deliberate simplification, sacrificing accuracy for implementation simplicity. The CTPAC method assumes that all circles share a common tangency, which is effective particularly when using adjacent circles but its performance is sensitive to measurement errors, especially in soils with non-linear behavior.\\
  
  Unlike p–q and CTPAC, which operate on fixed reference points, LSVD employs dynamic point selection: the set of virtual tangents changes with each iteration. This improves flexibility but increases computational complexity, particularly for non-linear envelopes where equation \ref{ends} may yield multiple solutions. In the power-law case, LSVD’s coefficients showed some variation but remained close to the tangency locations, suggesting convergence to local minima.\\
  
  Introducing controlled random noise into the input data effectively simulates measurement errors, enabling assessment of each method's robustness. This noise reveals how variability in soil strength propagates through the analysis. While increasing the number of samples generally improves the resolution of the failure envelope, three to six samples are typically sufficient for practical applications.\\
  
  Under noisy conditions, LSVD and CTPAC maintained precision in locating tangency points, and p-q consistently failed to capture tangency locations in several test cases.\\
  
  The absence of an exact analytical failure envelope precludes direct accuracy comparison. All three methods achieve high goodness-of-fit relative to their own reference points virtual tangents for LSVD, adjacent tangent points for CTPAC, and radii for p-q making the coefficient of determination ($R^2$) an unsuitable measure of tangency fidelity. Variability analysis and confidence interval coverage offer a more meaningful basis for comparison.\\
  
  In the linear case, LSVD can recover exact solutions when they exist; however, CTPAC is computationally faster as it requires only a single evaluation. For logarithmic and power-law envelopes. The p-q method struggled with linear and logarithmic envelopes but performed comparably to the other methods in the parabolic case. Importantly, finding an exact solution should not be considered a complete theoretical success, as the Mohr-Coulomb model itself is an approximation of soil behavior.\\
  
  Ultimately, method selection should balance computational cost, robustness to measurement noise, and the need for precise tangency localization. LSVD offers higher precision and adaptability across envelope forms but may require greater numerical effort. CTPAC is advantageous for linear cases when computational efficiency is critical and data quality is high, while p-q is best reserved for preliminary or low-precision applications.
\section{Conclusion}
  The Least Squares with Virtual Displacements explicitly use of tangency information results in higher precision compared to CTPAC and p–q, particularly under noisy conditions.\\
  
  The methods compare herein degrade significantly when measurement errors are introduced.\\
  
  Three to six samples provide an adequate balance between resolution, practicality and economy; additional samples would offer diminishing returns.\\
  
  The Least Squares with Virtual Displacements outperforms for linear, logarithmic, and power-law envelopes, while all methods perform similarly for parabolic forms.\\
  
  LSVD’s dynamic point selection enables broader applicability but increases computational complexity; CTPAC is more efficient for linear problems, and p–q remains the simplest but least reliable.\\
  
  LSVD provides a robust framework for determining failure envelopes under the Mohr–Coulomb criterion, effectively reducing uncertainty from methodological assumptions. While it may require more computational resources particularly for non-linear envelopes, due to the convergence of the solution and that some models are particular with their domain so it requires a constrained minimization, it offers superior robustness to noise and flexibility in adapting to different failure criteria, making it the preferred choice when precision is prioritized over computational simplicity. Even with static points, methods such as the power is not exempt of the non convexity problem, so the solution will depend on the initial guess of the optimization method.\\
  
  Drawing upon mathematical methods to get a soil's strength is finer and even quicker than a graphical method and is better to perform optimal designs.\\
  
  The linear LSVD is easy to implement in a spreadsheet (see appendix \ref{appendixe}) or coding it (see appendix \ref{appendixb}).\\
\bibliographystyle{ieeetr}
\bibliography{ref}

@article{trist,
	author = {Guo, Bao and Wang, Long and Li, Yizhe and Chen, Yan},
	year = {2020},
	month = {10},
	pages = {1-13},
	title = {Triaxial Strength Criteria in Mohr Stress Space for Intact Rocks},
	volume = {2020},
	journal = {Advances in Civil Engineering},
	doi = {10.1155/2020/8858363}}

@article{yuanming,
	title = {Stress-Strain Relationships and Nonlinear Mohr Strength Criteria of Frozen Sandy Clay},
	journal = {Soils and Foundations},
	volume = {50},
	number = {1},
	pages = {45-53},
	year = {2010},
	issn = {0038-0806},
	doi = {https://doi.org/10.3208/sandf.50.45},
	url = {https://www.sciencedirect.com/science/article/pii/S0038080620302201},
	author = {Lai Yuanming and Gao Zhihua and Zhang Shujuan and Chang Xiaoxiao}
}

@article{calib,
	title = {Calibration of Four Nonlinear Failure Envelopes from Triaxial Test Data and Influence of Nonlinearity on Geotechnical Computations},
	journal = {Geomaterials},
	number = {11},
	pages = {42-57},
	year = {2021},
	doi = {https://doi.org/10.4236/gm.2021.112003 },
	url = {10.4236/gm.2021.112003 },
	author = {Amaechi J. Anyaegbunamorcid and Fidelis O. Okafor},
}
\appendix

\section{Programs}
	\label{appendixb}
	The script to use LSVD can be found on the following github \href{https://github.com/xiaixue/lsvd}{repository}, it is written in Python. The required libraries are \texttt{numpy} and \texttt{scipy}, although \texttt{matplotlib} is also used to plot the results.
	
\section{Condition when $n=1$}
  \label{appendixc}
  If $n=1$ then the sums \textit{disappear} and the linear LSVD equation take the form
  $$
    0 = \beta_1(r^2 + \lambda^2) + \beta_0 \left( 
    \lambda - \dfrac{r\beta_1}{\sqrt{\beta_1^2+1}}
    \right)
    - \dfrac{2\beta_1^2+1}{\sqrt{\beta_1^2+1}} r\lambda
  $$
  where $\beta_0$ has also the form
  $$
    \beta_0 = r \sqrt{\beta_1^2+1}  - \beta_1 \lambda
  $$
  Thus, the equation simplified as
  $$
  0 = r\lambda \sqrt{\beta_1^2 + 1} + r\lambda \dfrac{\beta_1^2}{\sqrt{\beta_1^2+1}}
  - \dfrac{2\beta_1^2+1}{\sqrt{\beta_1^2+1}} r\lambda
  $$

  $$
  0 = r\lambda \left( \dfrac{(\beta_1^2+1)+\beta_1^2}{\sqrt{\beta_1^2 + 1}}\right) 
  - \dfrac{2\beta_1^2+1}{\sqrt{\beta_1^2+1}} r\lambda
  $$
  Which leads to $0=0$, therefore the when $n=1$ the equation \ref{sol2} will always sum 0, so any line coming from a $\beta_1$ and $\beta_0$ will generate zero error.
\section{Cohesionless condition verification}
  \label{appendixd}
  If $n=1$, the cohesionless condition is when $\beta_0$ is 0, so from equation \ref{sol1} we get $r\sqrt{\beta_1^2+1} - \beta_1 \lambda = 0$ which leads to the following quadratic equation $\left( 1 - \frac{r^2}{\lambda^2}\right) \beta_1^2 - \frac{r^2}{\lambda^2} = 0$.\\
  The solution to the quadratic equation is $\beta_1= \pm \left( 2-\frac{2r^2}{\lambda^2} \right) ^{-1} \sqrt{4 \frac{r^2}{\lambda^2} - 4 \frac{r^4}{\lambda^4}}$, by taking the positive part it leads to $\beta_1 = \left( 1 - \frac{r^2}{\lambda^2} \right) ^{-1} \frac{1}{\lambda} \sqrt{r^2 - \frac{r^4}{\lambda^2}}$. So it can be lead to the form.
  $$\beta_1=\lambda^{-1}\sqrt{r^2 - \frac{r^4}{\lambda^2}} \left( 1 - \frac{r^2}{\lambda^2} \right) ^{-1} \left( \frac{
    \sqrt{r^2 - \frac{r^4}{\lambda^2}}} {\sqrt{r^2 - \frac{r^4}{\lambda^2}}}\right) $$
  $$\beta_1= \frac{r}{\lambda \sqrt{1-\frac{r^2}{\lambda^2}}} \left( \frac{r}{\sqrt{r^2}} \right) = \frac{r^2}{\lambda \sqrt{r^2-\frac{r^4}{\lambda^2}}}$$
  Now, let $\hat{\tau}$ be the linear failure envelope and $\tau$ the Mohr circle equation. For a cohesionless soil, the shear strength is $\hat{\tau} = \left( \frac{d\tau (\prescript{\theta}{}{\sigma})}{d\sigma}\right)  \sigma$, where $\prescript{\theta}{}{\sigma}$ is the normal stress where it is tangent to the envelope. To know the $\prescript{\theta}{}{\sigma}$ the perpendicular function ought to be stablished such that $\hat{\tau}_{\perp}(\lambda) = 0$, so the perpendicular function is $\hat{\tau}_{\perp} = -\left( \frac{d\tau}{d\sigma}\right) ^{-1} \sigma + \left( \frac{d\tau}{d\sigma}\right) ^{-1} \lambda$. By equating the perpendicular with the failure envelope $\hat{\tau}(\prescript{\theta}{}{\sigma}) = \hat{\tau}_{\perp}(\prescript{\theta}{}{\sigma})$ it can be known $\prescript{\theta}{}{\sigma}$ by solving the equation.
  $$0= \frac{\prescript{\theta}{}{\sigma}^2 - \lambda \prescript{\theta}{}{\sigma}}{\sqrt{r^2 - (\prescript{\theta}{}{\sigma}- \lambda)^2}} + \sqrt{r^2 - (\prescript{\theta}{}{\sigma} - \lambda)^2}$$
  For that equation, the solution is $\prescript{\theta}{}{\sigma} = \frac{\lambda^2 - r^2}{\lambda}$. And that will lead to a tangent line $\beta_1 = \tan \varphi = \frac{-\prescript{\theta}{}{\sigma}+\lambda}{\sqrt{r^2-(\prescript{\theta}{}{\sigma}-\lambda)^2}}$
  After some algebra manipulation, $\beta_1$ can take the form $\frac{r^2}{\lambda\sqrt{r^2-\frac{r^4}{\lambda^2}}}$, the same as obtained from the virtual displacement.
\section{Example of the linear case}
  \label{appendixe}
  Here it is presented a worked example with the sample 1 from the data of table \ref{tabul}, it can be taken to a spreadsheet. According to equation \eqref{sol2} the terms are developed on table \ref{tabul1}.\\
    \begin{table*}[h!]
    \centering
    \caption{Data to calculate the linear failure envelope of sample 1. }
    \label{tabul1}
    \begin{tabular}{ c c c c c c c }
      \hline
      & $\sigma_3$ $[kg/cm^2]$ & $\sigma_1$ $[kg/cm^2]$& $r_i$ & $\lambda_i$ & $r_i^2+\lambda_i^2$ & $r_i \lambda_i$\\
      \hline
      1 &	0.5	& 2.211 & 0.8555	& 1.3555 & 2.5693 	& 1.1596 \\
      2 &	1	& 3.242 & 1.121	& 2.121 & 5.7553 	& 2.3776 \\
      3 &	1.5	& 3.909 & 1.2045	& 2.7045 & 8.7651 	& 3.2576 \\
      4 &	2	& 4.885 & 1.4425	& 3.4425 & 13.9316 & 4.9658 \\
      \hline
      $\sum$ &  &  &  &  & 31.02130 & 11.76065 \\
      \hline
    \end{tabular}
    \end{table*}
  The arithmetic means of the radii and the centers are the following:\\
  $$ \bar{r} = \frac{1}{4}\sum_{i=1}^{4} r_i = 1.155875 $$ 
  $$ \bar{\lambda} = \dfrac{1}{4}\sum_{i=1}^{4} \lambda_i = 2.405875 $$
  By setting an arbitrary starting $\beta_1 = 2$ then $
  \beta_0 = 1.155875 \sqrt{2^2+1} - 2 \times 2.405875 = -2.227134927$.\\
  From \eqref{sol2} we can treat it like
  $$
  0=\underbrace{\beta_1 \sum_{i} \left( \lambda_i^2 + r^2_i \right)}_{\xi_1}
  + \underbrace{n \beta_0 \left( \bar{\lambda} - \dfrac{\bar{r}\beta_1}{\sqrt{\beta_1^2+1}} \right)}_{\xi_2}
  \underbrace{- \dfrac{2\beta_1^2 + 1}{\sqrt{\beta_1^2+1}}\sum_{i} r_i\lambda_i } _{\xi_3}
  $$
  Then $\xi_1 = 62.04259$, $\xi_2 = -12.22277$, $\xi_3 = -47.33569$, so the sum is $2.484122346386$, which is not zero. Thereafter by solving the equation (using \textit{goal seek} on a spreadsheet) yields $\beta_1 = 0.281431718$. With that result is held $\xi_1 = 8.73038$, $\xi_2 = 4.383771649$, $\xi_3 = -13.11417$ that makes the equation $0.000025032518 \approx 0$. So the linear failure envelope is
  $$ \hat{\tau} =  \tan(15.718285839\deg) \sigma + 0.523688143$$
  
  \begin{figure}[!h]
  	\centering
  	\includegraphics[width= 400pt]{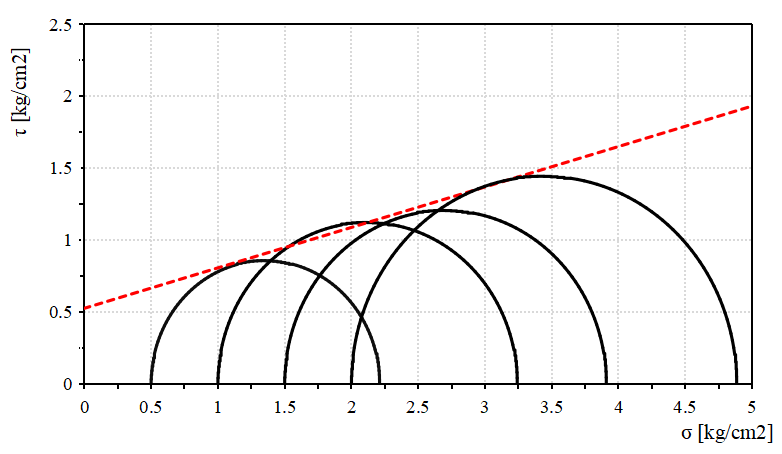}
  	\caption{Plot of computed failure envelope.}
  \end{figure}
\end{document}